\documentclass[pre,twocolumn,a4paper,noshowpacs]{revtex4}
\usepackage{amsmath}
\usepackage{amssymb}
\usepackage{amsfonts}
\usepackage{graphicx}
\usepackage{subfigure}
\usepackage{latexsym}
\usepackage{wasysym}

\begin{document}

\title{Critical behavior in a cross-situational lexicon learning scenario}

\author{P. F. C. Tilles and J. F.  Fontanari}

\affiliation{Instituto de F\'{\i}sica de S\~ao Carlos,
  Universidade de S\~ao Paulo,
  Caixa Postal 369, 13560-970 S\~ao Carlos, S\~ao Paulo, Brazil}

\begin{abstract}
The associationist account for early word-learning is based on the co-occurrence between objects and words. Here we examine the
performance of a simple associative learning algorithm for acquiring the referents of words in a cross-situational scenario affected by noise produced by out-of-context  words.
We find a critical value of the noise parameter $\gamma_c$  above which learning is impossible. We use finite-size scaling to show that the sharpness of the transition
persists across  a region of order $\tau^{-1/2}$  about $\gamma_c$, where $\tau$ is the number of learning trials, as well as to  obtain the  learning error (scaling function)
in the critical region. In addition,  we show that the distribution of durations of periods when the  learning error is zero is a power law  with exponent $-3/2$ at the critical point.
\end{abstract}

\maketitle

\section{Introduction}\label{sec:1} 

The problem of early word-learning   has been subject of philosophical 
controversy for centuries \cite{Bloom_00}. The  always visionary Augustine argued that the child makes the connections between words and their referents by understanding 
the referential intentions of others,  thus  anticipating the modern theory of mind in about fifteen centuries \cite{Adolphs_03}.  In the 17th century,  Locke's  empiricism 
supported the associationist viewpoint, which contends that the mechanism of word learning is 
sensitivity to covariation, i.e., if two events occur at the same time, they become associated. 

Here we  examine a radical  offshoot of the associationist approach to lexicon acquisition  termed cross-situational or observational learning \cite{Pinker_84},
which asserts  that the meaning of a word  can be determined by looking for something in common across all observed uses of that word 
\cite{Yu_12}.  In other words, learning takes place through the statistical sampling of the contexts in which a word appears. 

A scenario to describe the lexicon acquisition process should take into account the inherent ambiguity of the learning task (i.e.,  many
distinct objects may be associated to the same word) as well as the noisy effect of out-of-context words (i.e.,  the uttered  word may not refer to any object in the context).
Whereas the noiseless scenario has been explored in great detail in the literature \cite{Smith_06,Blythe_10,Tilles_12}, 
where it was shown that the learning error decreases exponentially with the number of learning trials, a systematic study of the effect of noise is
lacking. 

To remedy this deficiency, we modify the minimal model of noiseless cross-situational learning \cite{Smith_06,Blythe_10,Tilles_12} 
so as to include the effect of noise produced by
out-of-context words.  Using Monte Carlo simulations  and finite-size scaling  we identify and characterize
a critical phenomenon that separates  the asymptotic regime where the lexicon can be acquired  without errors from the regime where
learning is impossible. At the critical noise level,  we find that the duration of the periods with  zero error is distributed by a power-law
distribution.


\section{Cross-situational learning scenario}\label{sec:2}

We assume that there are $N$ objects,  $N$ words and a one-to-one mapping between words and objects.  At each learning event, $C$ 
objects are chosen  at random  without replacement from the fixed  list of $N$ objects and one of these objects is named according to the word-object mapping. 
The $C$ objects form the context which determines the interpretation of the  uttered  word  and the learner's task is to guess which of
the $C$ objects  that word refers to. This is then an ambiguous word-learning scenario in which there are multiple object candidates for any word. 
The parameter $C$ is a measure of the ambiguity of the learning task. In particular,  in the case $C=N$  the word-object mapping is not
learnable within a cross-situational scenario.
 
A learning episode  comprises a context  and a single target word.   In an uncorrupted learning episode, the context must exhibit the 
correct object (i.e., the object named by the target word according to the object-word mapping) plus $C-1$ distinct mismatching objects. 
 Noise is added to the  learning scenario by removing the correct object from the context, which will then  exhibit  $C$ mismatching objects.  Such  corrupted
and misguiding learning  episodes occur with probability $\gamma \in \left[0,1 \right]$. This type of noise is an integrant part of  any realistic learning situation,
arising usually from the  unwarranted  narrowing of the context by the learner.

To represent  the one-to-one object-word mapping we use the index $i=1,...,N$ to label the distinct objects and $h=1,...,N$ to label the 
distinct words. Then, without lack of generality, the correct mapping is defined by assigning  object $i=1$ to  word $h=1$, object $i=2$ to  word $h=2$ and so on. 
The problem faced by the learner is to determine the correct mapping  given a sequence of learning episodes. Next we will
describe a simple  (perhaps, the simplest) procedure  to accomplish this learning task.

\section{Associative learning model}\label{sec:3} 

We assume that learning is a change in the confidence with which the learner associates the target word $h$ to a given object $i$  and 
 represent this confidence by  a non-negative integer $p_{ih}$. Our associative accumulator learning procedure is described as follows.
Before  learning  all confidences are set to zero, i.e., $p_{ih} = 0$ for $i,h=1,...,N$, and whenever object $i^{\ast}$ appears in a context with target word $h^{\ast}$ the confidence $p_{i^{\ast} h^{\ast}}$ increases by one unit  \cite{Bush_55}. Hence, exactly  $C$ confidence values are updated at each learning trial. 

To determine which object corresponds to word $h$ the learner simply chooses the object index $i$ for which $p_{ih}$ is maximum. 
In the case of ties, the learner selects one object at random among those that maximize the confidence. From the definition of the correct word-object mapping, our  
learning algorithm achieves a perfect performance when $p_{hh} > p_{ih}$ for all $h$ and $i \neq h$. 

A critical feature of the accumulator model  is  that words are learned independently. This fact alone allows us to split the analysis of the vocabulary learning
task in two parts. The first and most important part is the problem of learning  the meaning (or the referent) of a {\it single} word.  Once this
is done, we can easily solve the problem of learning the $N$  words given their sampling frequencies \cite{Tilles_12}. Hence, in this work we will focus 
on the single-word learning problem only.

\section{Single-word learning}\label{sec:4} 

Accordingly, we consider the learning of a  single word, say word $h$,  which is then uttered at
all learning trials $\tau$. We define the single-word learning error  $\epsilon \left( \tau\right)$ for $\tau > 0$ as follows.
If $p_{hh} < p_{ih}$ for any  $i \neq h$ then $\epsilon = 1$, otherwise
if $p_{hh} = p_{ih}$ for $n$ values of $i \neq h$ then 
$\epsilon = n / \left(n +1\right)$  with $n=0, \ldots, N-1$.
At $\tau  =0$ all confidences are set to zero and so  $\epsilon = \left ( N - 1 \right )/N$.

In  the noiseless case ($\gamma = 0$) we have $p_{hh} \geq p_{ih}$ for all  $i \neq h$ since object $i=h$ is  always part of the context. 
So errors are due to ties $p_{hh} = p_{ih}, i \neq h $  only.
In fact, it can be shown analytically that in this case the average learning error vanishes  like $ \left [  \left ( C-1 \right ) / \left (N-1 \right )   \right ]^\tau $ 
for large $\tau$ \cite{Smith_06,Blythe_10,Tilles_12}. As expected, for $C=1$ we have  $\epsilon = 0$ at the first learning trial $\tau = 1$  already, but  more interestingly  
is that learning becomes faster with increasing $N$. This apparently 
counterintuitive result has a simple explanation: a  large list of objects to select from  actually decreases  the odds of choosing the same  confounding object  
during the learning trials, thus  reducing the number of ties.  However, this decrease is overcompensated by the sampling effect when we consider
the problem of learning the entire vocabulary and then learning slows down as $N$  increases, as expected \cite{Tilles_12}.

\begin{figure}[!t]
\centering
\includegraphics[width=1.0\linewidth]{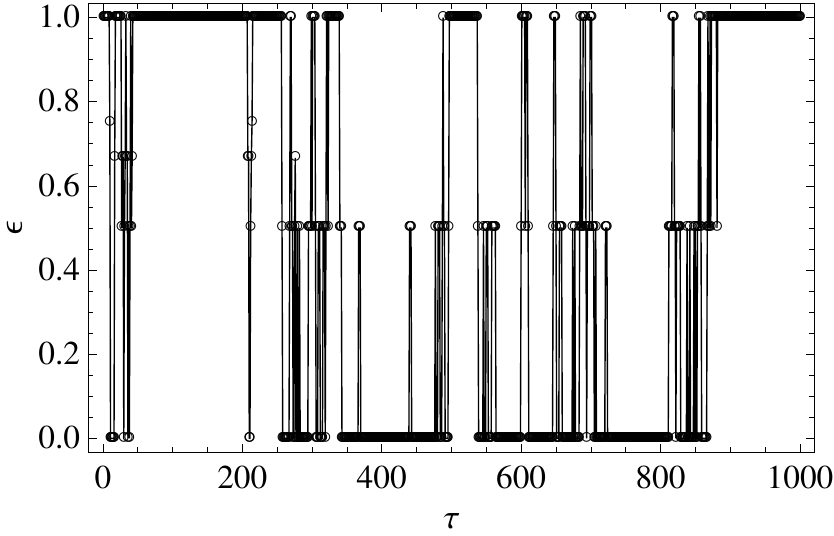}
\caption{Learning error against the number  of learning trials $\tau$ for a single sample  of the learning process using the  accumulator learning model.
The parameters are  $N=20$, $C=6$ and $\gamma =\gamma_c = 0.7$. 
The lines are guides to the eye.}
\label{fig:1}
\end{figure}

\begin{figure}[!t]
\centering
\includegraphics[width=1.0\linewidth]{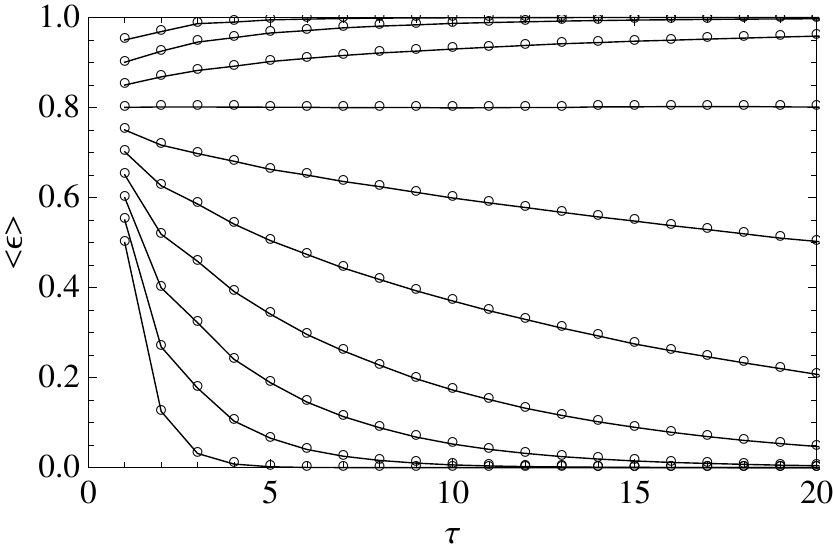}
\caption{Average learning error $\langle \epsilon \rangle$ as function of the number of learning trials for $N=5$, $C=2$ and (bottom to top) $\gamma = 0, 0.1, 0.2, ..., 0.9$. The critical value of the noise parameter is $\gamma_{c} = 0.6$ at which $ \langle \epsilon_{c} \rangle  = 0.8$. The symbols are the simulation results and the lines are guides to the eyes.}
\label{fig:2}
\end{figure}

In the case the contexts are corrupted by noise with a probability $\gamma$ an analytical approach is not  possible and we have to 
resort to simulations to study  the stochastic learning process. Figure \ref{fig:1} shows a typical evolution of the learning error at the critical
 noise level.  Although this figure reveals a rich stochastic 
dynamics,  it is rather uninformative from the learning perspective. In that sense,
the behavior of the average learning error $\langle \epsilon \rangle$, shown in Fig.\ \ref{fig:2},  is more relevant. For
a fixed $\tau$, this average is calculated using typically $10^6$ to $10^7$ realizations of the learning process. 

Figure \ref{fig:2} reveals that learning is possible  provided that the noise parameter does not exceed a certain threshold $\gamma_c$. More pointedly, 
in the asymptotic regime $\tau \rightarrow \infty$ we find that $\langle \epsilon \rangle \rightarrow 0$
for $\gamma < \gamma_c$ and that  $\langle \epsilon \rangle \rightarrow 1$ for $\gamma > \gamma_c$. The surprising finding is that at $\gamma = \gamma_c$, the average learning error becomes independent of $\tau \geq 0$.

There is
a simple  reasoning to determine $\gamma_c$ as well as the error $\langle \epsilon_c \rangle$ at this critical noise parameter.
First, we note that the borderline between learning and non-learning occurs when all $N$ objects are equally likely of being selected to compose 
the contexts.  We recall that this is exactly  the situation prior to learning and so we expect that
\begin{equation}\label{E2c}
\langle \epsilon_{c} \rangle= \epsilon \left( \tau=0\right)  = \frac{N-1}{N} .
\end{equation}
 Accordingly, $\gamma_c$  is determined  by equating the probability of selecting the correct object 
with the probability of selecting any given incorrect object to compose the context in a learning episode, i.e.,
\begin{equation}\label{G1c}
1 - \gamma_{c} = \left( 1 - \gamma_{c}\right)  \frac{C-1}{N-1} + \gamma_{c} \frac{ C}{N-1}, 
\end{equation}
from which we get 
\begin{equation}\label{G2c}
\gamma_{c} = 1 -\frac{C}{N}.
\end{equation}
These neat expressions for $\langle \epsilon_{c}  \rangle$ and $\gamma_c$ proved correct for a vast selection of values of $N$ and $C$, but we have no mathematical
proof of their validity, besides the  argument presented above.
However, we can perform a simple consistency check on these expressions as follows. 
The average learning error at the first trial is given by
\begin{equation}\label{E1c}
\langle \epsilon \left( \tau=1\right) \rangle  =  \left(1 - \gamma \right) \frac{C-1}{C} + \gamma
\end{equation}
and by setting $\gamma = \gamma_c$ we recover Eq.\ (\ref{E2c}) as it should be since $\langle \epsilon_{c} \rangle $ is independent of $\tau$ 
(see Fig.\ \ref{fig:2}).

\section{Finite-size scaling analysis}\label{sec:5}

\begin{figure}[!t]
\centering
\includegraphics[width=1.0\linewidth]{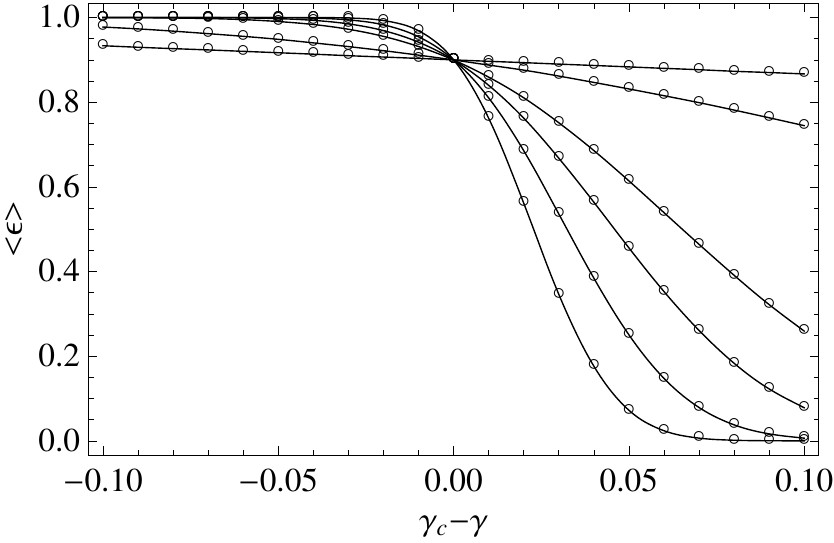}
\caption{Average learning error as function of the distance to the critical noise parameter for $N=10$ and $C=2$. The symbols are 
the simulation results for (top to bottom in the positive ordinate region) $\tau=1, 10, 100, 200, 400 $ and $800$.
The lines are  guides to the eyes.}
\label{fig:3}
\end{figure}

Considering the `size'  of the system as the number of learning trials $\tau$ we proceed now to examine the sharpness of the phase transition at
$\gamma_c$  using  finite-size scaling \cite{Privman_90}. This threshold phenomenon is best appreciated  in Fig. \ref{fig:3}, which  exhibits the dependence of the average 
learning error on the distance to the critical parameter for different values of $\tau$. As the number of trials $\tau$ increases, the difference between the regimes $\gamma < \gamma_{c}$ and
 $\gamma > \gamma_{c}$ becomes evident.  All curves intersect at 
$\gamma = \gamma_c$ for which the average error is a constant given by Eq.\ (\ref{E2c}).

\begin{figure}[!t]
\centering
\includegraphics[width=1.0\linewidth]{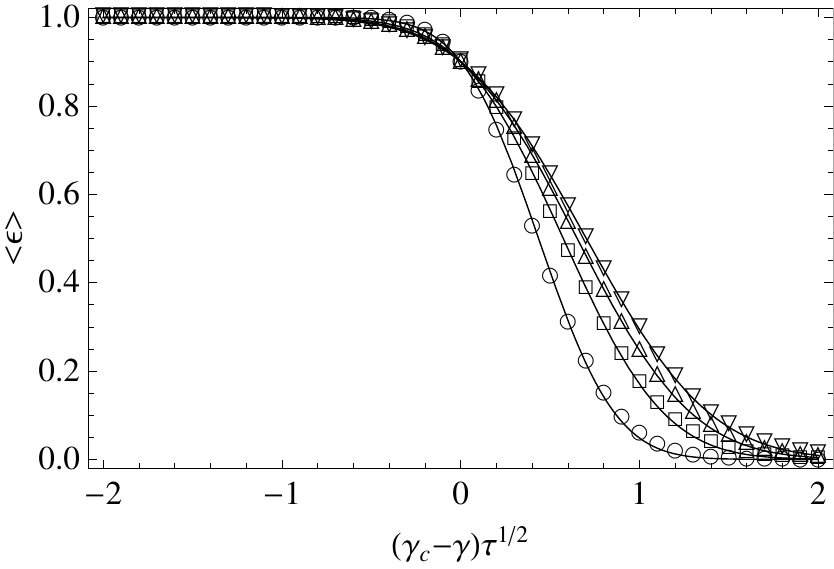}
\caption{Average learning error as function of the reduced variable $\left(\gamma_{c} -\gamma\right) \tau^{1/2}$ for $N=10$ and 
$C=1 (\bigcirc), 2 (\Box), 3 (\bigtriangleup)$ and $ 5 (\bigtriangledown) $. The symbols are the simulation results 
 and the lines are 
given by the scaling function (\ref{ansatz}) with the parameter $b$ obtained from the fitting of the data.}
\label{fig:4}
\end{figure}

The key insight is obtained when one considers the average learning error as a function of the
reduced variable $\left(\gamma_{c} -\gamma \right) \tau^{1/2}$, as exhibited in Fig.\ \ref{fig:4}.
Use of this reduced variable produces the collapse of the data for different $\tau$ into a single  scaling function, which depends on the values of $N$ and $C$ only. 
As illustrated in the figure, the data is fitted very well by the functional form
\begin{equation}\label{ansatz}
\langle \epsilon  \rangle  = \frac{1}{2} \textrm{erfc}\left[ a\left( N \right) + b\left(N,C\right) \left(\gamma_{c} -\gamma\right) \tau^{1/2} \right], 
\end{equation}
which has a single  fitting parameter,  $b \left(N,C\right) $. The parameter $a\left( N \right)$  is obtained by setting $\gamma = \gamma_{c}$ and then using the
expression of $\langle \epsilon_c \rangle$, given by Eq.\ (\ref{E2c}). The final result is 
\begin{equation}\label{a}
a \left( N\right) = \textrm{erfc}^{-1}\left[\frac{2\left(N-1\right)}{N}\right],  
\end{equation}
where $\textrm{erfc}^{-1} \left ( x \right ) $  stands for the inverse complementary error function. We note that $a \left ( 2 \right ) =  0$ and
$a \left ( N \right ) < 0$ for $N > 2$. 

\begin{figure}[!t]
\centering
\includegraphics[width=1.0\linewidth]{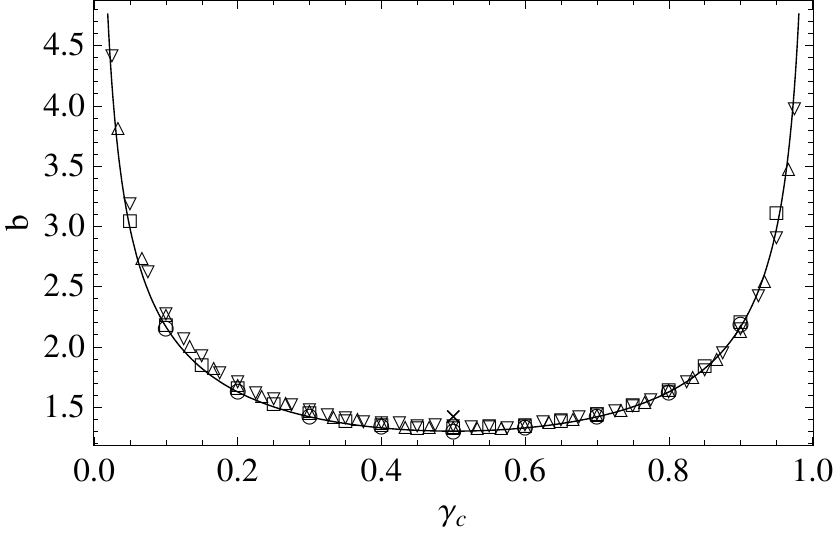}
\caption{Dependence of the fitting parameter  $b$ on the ratio $ \gamma_c$ for $N=2 (\times)$,  $N=10 (\bigcirc)$, $N=20 (\Box)$, $N=30 (\bigtriangleup)$ and $N=40
(\bigtriangledown)$. 
The solid line is  given by  Eq.\ (\ref{b}). }
\label{fig:5}
\end{figure}

We can get some insight on the fitting parameter $b \left(N,C\right) $ by calculating explicitly the
average learning error for $N=2$ and $C=1$.  In the limit
$\tau \to \infty$ and $ \gamma \to \gamma_c = 1/2$ such that  $\tau^{1/2} \left (\gamma - \gamma_c \right ) $ is finite,
we find to the leading order
\begin{equation}\label{A2}
\langle \epsilon \rangle  \sim
\frac{1}{2} \mbox{erfc} \left [  \frac{  \tau^{1/2} \left (\gamma_c - \gamma\right ) }{ \left [ 2 \gamma_c \left( 1 - \gamma_c \right) \right ]^{1/2}}  \right ] .
\end{equation}
Hence we assume that $b \left(N,C\right) = b \left( \gamma_{c}\right)$  and plot this fitting parameter in   Fig.\ \ref{fig:5} for a large selection of  
values of $N$ and $C$.  More pointedly, for each value of $N$ (represented by different symbols in the figure) we vary $C$ from $1$
to $N-1$ to  obtain scaling functions as those shown in Fig.\ \ref{fig:4}. Then these functions are fitted using Eq.\ (\ref{ansatz}) in order to
determine the fitting parameter $b$. For $N >4$ the data is fitted very well by the function
\begin{equation}\label{b}
 b \left( \gamma_{c}\right) = \frac{b'}{ \left [  \gamma_c \left( 1 - \gamma_c \right) \right ]^{1/2}}
\end{equation}
with  $b' = 0.65$. Note that for $N=2$ we have $b'= 1/\sqrt{2} \approx 0.71$.  

Figure \ref{fig:5} reveals a most interesting symmetry: for fixed $N$ the average  learning error when plotted against the reduced variable 
$\left ( \gamma_c - \gamma \right ) \tau^{1/2}$  is invariant to the change $ C \to N - C$ which implies
$\gamma_c  \to 1 - \gamma_c$. In particular, in Fig.\ \ref{fig:4} the results for $C=9$ are identical to those displayed for
$C=1$, the results for $C=8$ to those for $C=2$ and so on. However, we must note that this symmetry  is exact only in the limits $ \tau \to \infty$  and
$\gamma \to \gamma_c$.

For an infinitely large lexicon, $N \to \infty$,  we have $ a \left( N\right) \sim - \ln^{1/2} N $ and so
$\langle \epsilon \rangle \to 1$ if the context size $C$ grows linearly with $N$ (i.e., $\gamma_c$ is nonzero), but
$\langle \epsilon \rangle \to 0$ if $C$ remains finite since in this case $b \sim N^{1/2}$ diverges faster than $a \left( N\right)$.

\section{Statistics of stasis}\label{sec:6}

A distinctive feature of the learning process revealed by Fig.\ \ref{fig:1} is the existence of long periods when the learning error stands at  zero value,
i.e., $p_{hh} > p_{ih}$ for all objects $i \neq h$. These periods or stases  are characterized by repeated additions of credence units to the 
confidence values $p_{ih}$ and they  end when   one (or more)  of the $N-1$ confidences $p_{ih}, i \neq h,$ equals $p_{hh}$.

\begin{figure}[!t]
\centering
\includegraphics[width=1.0\linewidth]{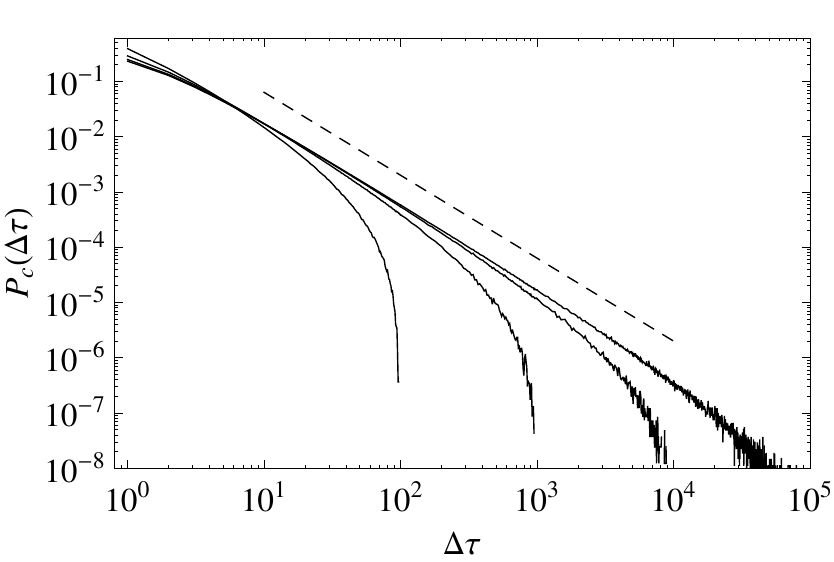}
\caption{ Distribution of stases for $N=20$, $C=6$, $\gamma = \gamma_c = 0.7$, and (bottom to top) $\tau_0 = 10^3, 10^4$ and $10^5$. The slope of the  straight line is $-3/2$.  }
\label{fig:6}
\end{figure}

We begin the analysis of  the distribution $P_c \left (  \Delta \tau \right ) $  of the  durations $\Delta \tau$ of the stases at the critical parameter $\gamma_c$ by
showing in Fig.\ \ref{fig:6} how  the total  number of learning trials  $\tau_0$ (basically a cutoff time) affects this distribution. The rescaling $\tau_0^{3/2} P_c \left (  \Delta \tau / \tau_0 \right ) $
makes the results essentially independent of the cutoff parameter $\tau_0$  provided  $\Delta \tau / \tau_0$  is not too small (data not shown). 
The curves exhibit a clear power law behavior with exponent $-3/2$, which  is the mean-field exponent for the size  of avalanches in self-organized critical models
\cite{Munoz_99}. 

In addition,  we find that away from the critical point the distribution  $P\left (  \Delta \tau \right ) $    is exponential and that the average duration of the stases diverges like 
$\langle \Delta \tau \rangle \sim \mid \gamma_c - \gamma \mid^{-1}$
as $\gamma \to \gamma_c$. 

As expected, these mean-field critical exponents are robust to changes in the model parameters $N$ and $C$.  In fact, for $N=2$ and $C=1$
the distribution $P \left (  \Delta \tau \right ) $ can be easily calculated analytically for any value of $\gamma$ since this is the classical ruin problem in which a gambler with initial
capital $z=1$ plays against an infinitely rich adversary. The results for the duration of the game $\Delta \tau $ are simply
$ P_c \left (  \Delta \tau \right ) \approx \left ( 2/\pi \right )^{1/2}  \left (  \Delta \tau \right )^{-3/2}$ and $\langle \Delta \tau \rangle = \left (1/2 \right ) \mid \gamma_c - \gamma \mid^{-1}$
(see Chapter XIV of \cite{Feller_68}).

\begin{figure}[!t]
\centering
\includegraphics[width=1.0\linewidth]{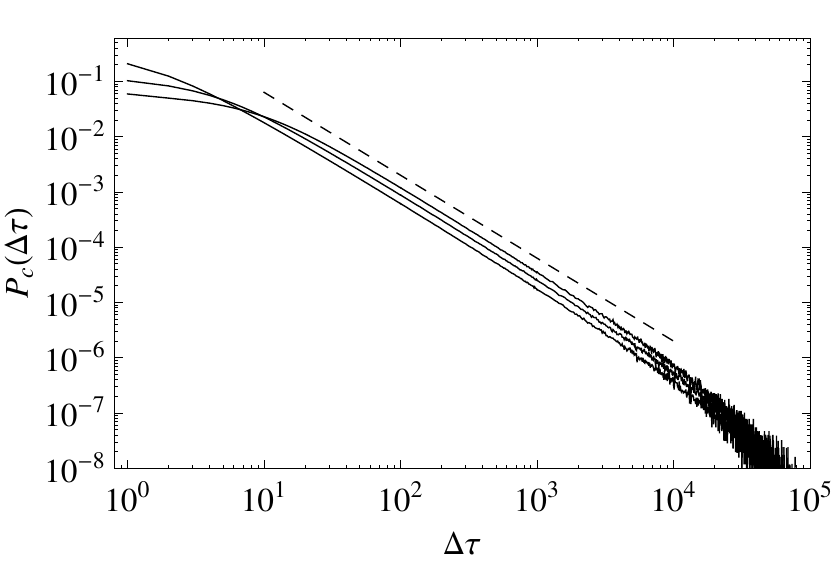}
\caption{ Distribution of stases for $N=20$, $\tau_0 = 10^5$ and (bottom to top at $\Delta \tau = 1$) $C=1,2,5 $. The slope of the  straight line is $-3/2$.  }
\label{fig:7}
\end{figure}

Changes in  the number of objects  $N$  have no significant influence  on  $P_c \left (  \Delta \tau \right )$ whereas  changes in the context size  $C$  produce
a shift on the distribution, without affecting the power-law exponent, as illustrated in Fig.\ \ref{fig:7}. In fact, increase of  $C$  increases the frequency of short stases and,  consequently,
reduces the frequency of  long ones.  This is expected since the larger the context size, the greater the number of mismatching objects that have their
confidences updated,  and so the greater the odds of occurrence  of the jump condition $p_{ih} \geq p_{hh}$ for some object $i  \neq h$.

%

%

Finally, we note that although we have focused on the periods of the learning process  when the error learning is $0$, the very same conclusions  hold for
the periods when the learning error is $1$.

\section{Conclusion}\label{sec:7}

The view of language as a collective phenomenon arising out of local social 
interactions has prompted its modeling and investigation  through statistical physics  concepts and tools\cite{Loreto_07}.  
Words have been likened to genes and their evolution studied
within a population genetics framework \cite{FP_04,Baxter_06}, whereas the competition between whole languages 
has been considered using population dynamics models  \cite{Abrams_03,Mira_05,Schulze_08}.  The study of the  bootstrap of a common lexicon
among a large population of individuals has revealed a sharp phase transition towards shared conventions \cite{Baronchelli_06} as well as an unexpected
connection with random occupancy problems in the case  only two individuals interact but the lexicon size is very large\cite{Fontanari_11}.

The problem of acquiring,  rather than bootstrapping, a fixed lexicon from observational learning is relevant to developmental psychology 
since it allows a quantitative appraisal of  the associationist hypothesis on  early-word learning \cite{Bloom_00}. In particular, we show that
the  utterance of out-of-context words may result in severe limitations to learning,  depending on the  ratio  $C/N$ between the number  of 
objects presented to the learner at a learning trial and the total number of objects. If this ratio is small (i.e., $\gamma_c$ is close to 1) then
this noisy effect is largely irrelevant  and the  lexicon can quickly be learned  to perfection. However, for large values of this ratio (i.e., $\gamma_c$ is close to 0)  
learning becomes impossible  regardless of the number of trials $\tau$.
Finite-size scaling shows that the threshold phenomenon  persists  across  a region of
size  $\tau^{-1/2}$  around $\gamma_c$  and  offers the explicit functional form of the learning error in this region.

The simplicity of our associative learning algorithm allowed us to consider  the learning of the distinct words as independent
stochastic   processes. Interactions between words, such as the  mutual exclusivity constraint  that
instructs children to associate novel words to unnamed objects \cite{Bloom_00}, are well-established in developmental
psychology and it would be interesting to see whether and how they alter  the characteristics of the critical phenomenon reported here.

\section*{Acknowledgments}

This research was supported by The Southern Office of
Aerospace Research and Development (SOARD), Grant No.
FA9550-10-1-0006, and Conselho Nacional de Desenvolvimento
Cient\'{\i}fico e Tecnol\'ogico (CNPq). P.F.C.T. was supported by  
Funda\c{c}\~ao de Amparo \`a Pesquisa do Estado de S\~ao Paulo 
(FAPESP).

\end{document}